\documentclass{desyprocA4}
\newcommand\pubdate{\today}
\def\eslt{E_T^{\rm miss}}

\def\esl{\not\!\!{E}}

\def\to{\rightarrow}

\def\bi{\begin{itemize}}
 \def\ei{\end{itemize}}

\def\c1p{C1^\prime}
\def\msq3{\overline{m}_{\tilde{q}}(3)}
\def\ta{\tilde a}

\def\ta{\tilde a}

\def\tst{\tilde t}

\def\tg{\tilde g}

\def\be{\begin{equation}}  
\def\ee{\end{equation}}  
\def\bea{\begin{eqnarray}}  
\def\eea{\end{eqnarray}}  
\def\tw{\widetilde W}

\def\tz{\widetilde Z}

\def\beq{\begin{equation}}
\def\eeq#1{\label{#1}\end{equation}}
\def\eeqn{\end{equation}}

\newenvironment{Eqnarray}%
   {\arraycolsep 0.14em\begin{eqnarray}}{\end{eqnarray}}
\def\beqa{\begin{Eqnarray}}
\def\eeqa#1{\label{#1}\end{Eqnarray}}
\def\eeqan{\end{Eqnarray}}

\begin{document}
\title{Leaving no stone unturned
in the hunt for \\SUSY naturalness: A Snowmass whitepaper}

\author{{\slshape H. Baer$^1$, V. Barger$^2$, P.~Huang$^2$, D.
    Mickelson$^1$, A.~Mustafayev$^3$, W. Sreethawong$^4$ and X. Tata$^3$}\\
$^1$Dept. of Physics and Astronomy, University of Oklahoma, Norman, OK 73019, USA\\ 
$^2$Dept. of Physics, University of Wisconsin, Madison, WI 53706, USA\\
$^3$Dept. of Physics and Astronomy, University of Hawaii at Manoa,
    Honolulu, HI 96822, USA\\
$^4$School of Physics, Suranaree University of Technology, Nakhon
    Ratchasima 3000, Thailand}


\maketitle

\pubdate

\begin{abstract}

Imposing electroweak scale naturalness constraints (low $\Delta_{EW}$) on
SUSY models leads to mass spectra characterized by light
higgsinos $\sim 100-300$~GeV, highly mixed top-squarks and
gluinos at the $1-5$~TeV scale and allows for $m_h \sim 125$~GeV. 
First and second generation squarks can easily live at the 5-20~TeV scale, 
thus providing at least a partial solution to the SUSY flavor/CP problems.  
For such models at the LHC, gluino pair production is followed by cascade decays to $t$-
and $b$-quark rich final states along with multileptons.  The reach of
LHC14 with 300~fb$^{-1}$ is computed to be around $m_{\tg}\simeq 1.8$
TeV. However, the small magnitude of the $\mu$-parameter -- 
a necessary condition for naturalness -- leads to
 a unique hadronically quite same-sign diboson ($W^\pm W^\pm$) signature from wino pair production. 
In low $\Delta_{EW}$ models with unified gaugino masses, 
this signature yields a somewhat higher reach up to $m_{\tg}\sim 2.1$~TeV. 
The smallness of $|\mu|$ implies that the {\it ILC should be a higgsino factory} in addition to a Higgs factory, 
and a {\it complete} search for SUSY naturalness seems possible for $\sqrt{s}\sim 600$~GeV. 
Since a thermal under-abundance of higgsino-like WIMP dark matter (DM) is expected,
there is ample room for an axion DM contribution.
A thorough search for higgsino-like WIMPs can be made by next generation WIMP detectors,
such as those with ton-scale noble liquid targets.
\end{abstract}

%

\section{Introduction}

In previous studies \cite{ltr,rns} and a companion contribution
Ref.~\cite{white1}, we have argued that {\it a necessary condition for
the naturalness} of SUSY models is the requirement that there are no
large cancellations in the familiar one-loop effective potential
minimization condition
\be \frac{M_Z^2}{2} =
\frac{m_{H_d}^2 + \Sigma_d^d -
(m_{H_u}^2+\Sigma_u^u)\tan^2\beta}{\tan^2\beta -1} -\mu^2 \;.
\label{eq:loopmin}
\ee 
Here, Eq.~(\ref{eq:loopmin}) is implemented as a {\it weak scale
relation}, even for SUSY theories purporting to be valid all the way up
to scales as high as $M_{\rm GUT}-M_{P}$.  The quantities $\Sigma_u^u$
and $\Sigma_d^d$ are the one-loop corrections arising from loops of
particles and their superpartners that couple directly to the Higgs
doublets.  The rationale for this, which has been presented in
Ref.~\cite{ltr,rns,white1}, led us to introduce the quantity
$\Delta_{\rm EW}$ which is determined essentially by the physical
super-partner spectrum. Spectra that lead to low values of $\Delta_{\rm
EW}$ offer the possibility of being derivable from natural theories.
Here, we summarize the phenomenological consequences of low $\Delta_{\rm
EW}$ models for LHC, ILC and dark matter searches.

The naturalness requirement of no large, uncorrelated contributions to $M_Z$ 
on the right-hand-side of Eq.~(\ref{eq:loopmin}) implies the following.
\begin{enumerate}
\item The magnitude of the weak scale $\mu$ parameter is not too far
from $M_Z$: $|\mu |\sim 100-300$~GeV to enjoy better than $3\%$
electroweak fine-tuning (lower $|\mu|$ gives less fine tuning).
\item The soft term $m_{H_u}^2$ is driven to small negative values
  with $|m_{H_u}^2|\sim 100-300$~GeV. This occurs in the FP region of
  mSUGRA/CMSSM but can occur at any $m_0,\ m_{1/2}$ values in less
  constrained models such as the two-extra-parameter non-universal Higgs model (NUHM2). 
\item The top squarks are at the few-TeV scale but highly mixed.
The large mixing suppresses the contributions $\Sigma_u^u(\tst_1,\tst_2)$ to $\Delta_{EW}$
  whilst raising $m_h$ up to the 125~GeV level\cite{ltr}.
\end{enumerate}

SUSY models which generate these conditions will automatically produce
$Z$ and $h$ masses around the 100~GeV scale while respecting LHC
constraints on $m_h$ and on sparticle masses, thus 
accommodating a Little Hierarchy (which in this case is no Problem). 
In SUSY models which are valid up to some high scale $\Lambda\sim M_{\rm GUT}$ or $M_P$, the soft term
$m_{H_u}^2$ can be radiatively driven to values $\sim -M_Z^2$; this
class of models is called {\it radiatively-driven natural SUSY}, or
RNS.  RNS with low electroweak fine-tuning $\Delta_{EW}$ can be realized
within the two-parameter non-universal Higgs model (NUHM2)\cite{nuhm2},
but not in more constrained models such as mSUGRA/CMSSM.\footnote{In
mSUGRA/CMSSM, $\mu\sim M_Z$ can be generated in the HB/FP region.
However, in that region the top-squarks are so heavy that the
$\Sigma_u^u$ radiative corrections dominate $\Delta_{EW}$ and lead to
fine-tuning typically at the 0.3\% level or worse.}

\begin{figure}[htb]
  \begin{center}
\includegraphics[width=0.6\textwidth]{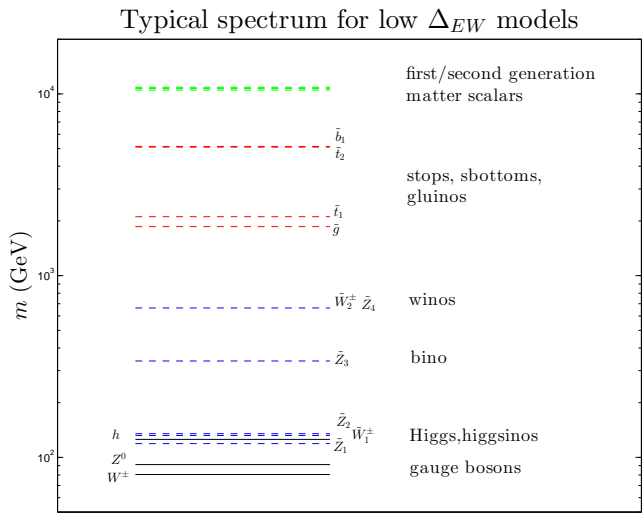}
  \end{center}
  \caption{Typical mass spectrum from low $\Delta_{EW}$ models}
\label{fig:spectra}
\end{figure}

RNS spectra are characterized by the following features, with a typical
spectrum shown in Fig.~\ref{fig:spectra}.
\begin{itemize}
\item Four light higgsinos $\tz_1$, $\tz_2$ and $\tw_1^\pm$ with mass
  $\sim 100-300$~GeV.  In fact, the lighter end of this range (closer
  to $M_Z$) is preferred by naturalness since $\Delta_{EW}\gtrsim |\mu^2|/(M_Z^2/2)$.
\item Highly mixed top- and bottom-squarks and also gluinos in the $1-5$ ~TeV range 
(this is significantly heavier than the range predicted by earlier natural SUSY models\cite{kn,pap}).
\item If gaugino mass unification is imposed, then $\tz_3$ will be bino-like at $\sim 0.2-0.8$~TeV
and $\tz_4$ and $\tw_2$ will be wino-like at $\sim 0.4-1.6$~TeV.
\item First/second generation squarks and sleptons may be at the $5-30$~TeV range, thus providing at
least a partial decoupling solution to the SUSY flavor/CP problems.
\end{itemize}

%

\section{RNS at LHC}

The phenomenological consequences of RNS models at LHC have been
explored in Ref's~\cite{lhcltr} and \cite{lhc}.  A brief summary of
highlights includes the following.
\begin{itemize}
\item Gluino pair production can be substantial at LHC14 for
  $m_{\tg}\sim 1-2$~TeV. In RNS models, gluinos cascade decay
  dominantly to the three-body modes $t\bar{t}\tz_i$ and
  $t\bar{b}\tw_i$, giving rise to the usual multilepton+multijet+$\eslt$ final states. 
  The LHC reach for gluino
  pair production at LHC14 with 300~fb$^{-1}$ is estimated to be
  $m_{\tg}\sim 1.8$~TeV~\cite{lhc,bblt}.

\item A qualitatively new SUSY signal for models with light higgsinos
  emerges: same-sign diboson (SSdB) production accompanied by modest
  jet activity. This signal arises from wino pair production
  $pp\to\tw_2^\pm\tz_4$ followed by $\tw_2\to W\tz_{1,2}$ and $\tz_4\to
  W^\pm\tw_1^\mp$.  Half the time one arrives at the SSdB final
  state. The very compressed higgsino spectrum means the decay products
  of $\tw_1$ and $\tz_2$ are very soft, so that accompanying hadronic
  activity emerges mainly from initial state QCD radiation. After cuts,
  the dominant background is found to be $Wt\bar{t}$ production. The
  reach of LHC14 with 300~fb$^{-1}$ is to $m_{\tg}\sim 2.1$~TeV
  \cite{lhcltr}, somewhat beyond the reach found from gluino pair
  production if the wino and gluino are related by gaugino mass
  unification. We emphasize that the SSdB signal from wino pair
  production is present in all models where winos are accessible and
  where higgsinos are light. As seen from Fig.~\ref{fig:reach},
  this signal can access $M_2\simeq 0.8m_{1/2}$ values as large as
  800~GeV at luminosity upgrades of the LHC. Moreover, in low $|\mu|$
  models, this signal offers a larger reach than the $gauginos\to WZ\to 3\ell +\eslt$
  trilepton signal.

\begin{figure}[htb]
  \begin{center}
\includegraphics[width=0.6\textwidth]{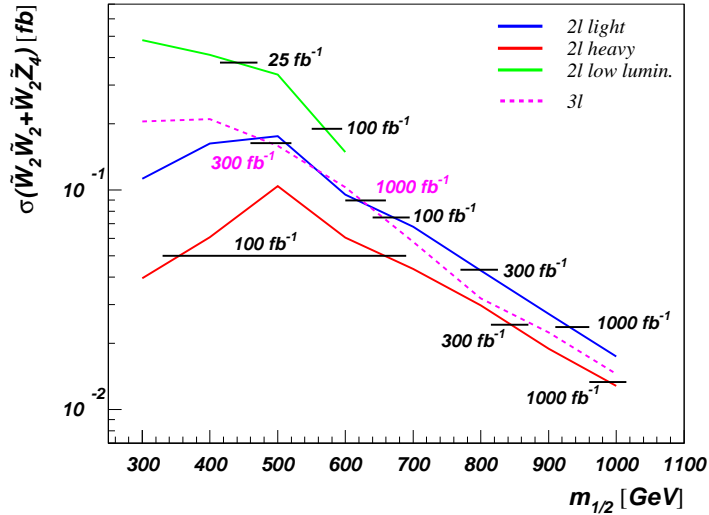}
  \end{center}
  \caption{Multilepton cross sections from wino pair production
  $\tw^\pm_2 \tz_4$ and $\tw^+_2\tw^-_2$ after cuts vs. $m_{1/2}$ along the
  RNS model line.  The reach of LHC14 is indicated for various
  integrated luminosity choices.  Solid curves represent the SSdB 
  channel for several set of cuts optimized for light (blue)
  and heavy (red) winos, and for low integrated luminosity (green):
  see Ref.~\cite{lhcltr}. The corresponding situation for 
  the $gauginos\to WZ\to 3\ell +\eslt$ channel is represented by the
  dashed magenta
  curve.}
\label{fig:reach}
\end{figure}

\item The reaction $pp\to\tw_2\tz_4$ and $\tw_2^+\tw_2^-$ gives rise to
  $WZ$ final states\cite{wz} which can be tagged via the much-studied
  trilepton signature. The reach of LHC14 with 300~fb$^{-1}$
  (1000~fb$^{-1}$) in this channel is to $m_{\tg}\sim 1.3$~TeV
  (1.65~TeV), well below the SSdB reach as already noted.

\item The wino pair production reactions also lead to an observable rate
  for $4\ell+\eslt$ events for integrated luminosity exceeding
  100~fb$^{-1}$. In fact, this channel yields a slightly better reach
  than via trilepton events just discussed, though not as high as via
  the SSdB search.

\item The reaction $pp\to\tw_1\tz_2$ gives rise to soft
  trilepton$+\eslt$ events.  These may be visible for the lower range
  of $m_{1/2}$ where the $\tz_2$ is mixed higgsino-bino with a larger
  mass gap $m_{\tz_2}-m_{\tz_1}\sim 30-40$~GeV. As $m_{1/2}$ (or $M_2$)
  increases, the mass gap drops to the 10-20~GeV range and the OS/SF
  dilepton from $\tz_2\to\ell^+\ell^-\tz_1$ decay gets lost under the
  $W^*\gamma^*$ background.
\end{itemize}

\section{RNS at ILC}

The key prediction of RNS models and SUSY naturalness is the existence
of four light higgsinos $\tz_1$, $\tz_2$ and $\tw_1^\pm$ with mass
$\sim |\mu|\sim 100-300$~GeV, the lower the better.  Thus, we would
expect the proposed International Linear $e^+e^-$ Collider (ILC) --
operating with $\sqrt{s}>2|\mu |$ -- to be a {\it higgsino factory} in
addition to a Higgs factory.  The main production reactions include
\begin{itemize}
\item $e^+e^-\to \tw_1^+\tw_1^-$ and
$e^+e^-\to \tz_1\tz_2$, $\tz_2\tz_2$.
\end{itemize}
The $\tw_1$ decays dominantly through $W^*$ into $f\bar{f}'\tz_1$ (where $f$ and $f'$ are SM fermions)
and $\tz_2\to f\bar{f}\tz_1$. 
Production cross sections for these reactions vs. $\mu$ for $\sqrt{s}=500$~GeV
and vs. beam polarization $P_L(e^-)$ have been shown in Ref.~\cite{bbh}.
For $e^+e^-$ collisions at $\sqrt{s}=500$ GeV and not too close to threshold, 
$\sigma(\tw_1^+\tw_1^- )\sim 400$ fb and $\sigma(\tz_1\tz_2 )\sim 150$ fb.

\begin{figure}[htb]
  \begin{center}
\includegraphics[width=0.6\textwidth]{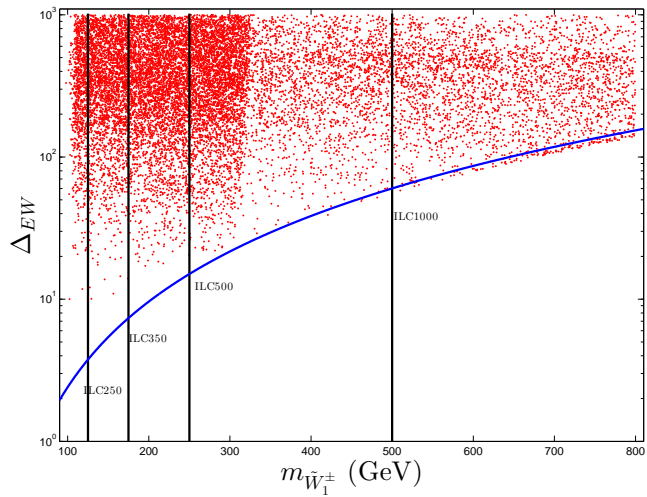}
  \end{center}
  \caption{ A plot of $\Delta_{EW}$ vs. $m_{\tw_1}$ from a scan of NUHM2
models.  We also show the projected reach of ILC with $\sqrt{s}=250,\
350,\ 500$ and 1000~GeV.}
\label{fig:ilc}
\end{figure}

Detailed studies of mixed higgsino-bino states from HB/FP SUSY with a
mass gap of $\sim 40$~GeV have been
presented in Ref's~\cite{bbkt,bkt}.
The reaction $e^+e^-\to\tz_1\tz_2\to\ell^+\ell^-+\esl$
should easily yield the $m_{\tz_2}-m_{\tz_1}$ mass gap to sub-GeV
precision.  By examining $e^+e^-\to\tw_1^+\tw_1^-\to
(\ell\nu_\ell\tz_1)+(q\bar{q}'\tz_1)$ states, the $m_{\tw_1}$ and
$m_{\tz_1}$ masses could be measured to better than $\sim 10\%$
assuming just 100~fb$^{-1}$ of integrated luminosity. Then the weak
scale Lagrangian parameters $M_2$ and $\mu$ could be determined to
20\% and 10\% precision, respectively.   
%
A detailed study of higgsino pair production for the much more
challenging case of the Brummer-Buchmueller (BB) model\cite{bb} where
the $\tw_1 -\tz_1$ mass gap is at the 1~GeV level\cite{jenny} has been
undertaken. Based on the early results of this analysis, together with
those from the earlier analyses mentioned, we are optimistic that the
RNS higgsino signal will be detectable at the ILC if $\sqrt{s} >
2|\mu|$.   A dedicated
analysis should, however, be carried out to confirm this.
Fig.~\ref{fig:ilc} shows that ILC500 (ILC1000) will decisively be able
to probe $\Delta_{\rm EW}$ values up to 15 (60) in this channel. 

%

\section{Direct/indirect detection of higgsino-like WIMPs from RNS}

The calculated thermal relic density of higgsino-like WIMPs from RNS has
been calculated in Ref's~\cite{rns,bbm}, and is typically found to be
$\Omega_{\tilde h}^{TP}h^2\sim 0.007-0.01$, {\it i.e.}  a factor 10-15
below measured values. The authors of Ref.~\cite{rns,bbm} suggest a
cosmology with mixed axion/higgsino dark matter (two dark matter
particles, an axion and a higgsino-like neutralino)\cite{ckls}. In such
a cosmology, thermal production of axinos $\ta$ in the early universe
followed by $\ta\to g\tg,\ \gamma\tz_i$ leads to additional neutralino
production. In the case where axinos are sufficiently produced, their
decays may lead to neutralino re-annihilation at temperatures below
freeze-out, which also augments the neutralino abundance. In addition,
coherent-oscillation production of saxions $s$ at high PQ scale
$f_a>10^{12}$~GeV followed by saxion decays to SUSY particles can also
augment the neutralino abundance. Late saxion decay to primarily SM
particles can result in entropy dilution of all relics (including
axions) present at the time of decay, so long as BBN constraints are
respected. The upshot is that, depending on additional Peccei-Quinn
parameters, either the higgsino-like neutralino or the axion can
dominate the dark matter abundance, or they may co-exist with comparable
abundances, leading to possible detection of both an axion and a
WIMP. In the case of the axion, it is straightforward to find acceptable CDM
densities with PQ scale $f_a\sim 10^{12}-10^{16}$~GeV, far beyond the
usual acceptable range from non-SUSY axion theories.

In the case of mixed axion-WIMP dark matter, the local WIMP abundance
might be well below the commonly accepted local abundance
$\rho_{loc}\sim 0.3\ GeV/cm^3$. Thus, current limits from experiments
like Xe-100 and CDMS should be scaled up by a factor $\xi\equiv
\Omega_{\tilde h}^{TP}h^2/0.12$ to be applicable. In the RNS model,
the gauginos also cannot be too light so that the neutralino always
has a substantial gaugino component even though it is primarily
higgsino.  This means that spin-independent direct detection rates
$\sigma_{SI}(\tz_1p)$ are never too small.  Even accounting for the
local scaling factor $\xi$, it is found in Ref.~\cite{bbm} that
ton-scale noble liquid detectors such as Xe-1-ton should completely
probe the model parameter space. One caveat is that if saxions give
rise to huge entropy dilution after freeze-out while avoiding constraints from
dark radiation and BBN, then the local
abundance may be even lower than the assumed freeze-out value, and the
dark matter would be highly axion dominated.

\section{Conclusions}

The main implication of natural SUSY models is that there should exist
four light physical higgsino-like states $\tz_1$, $\tz_2$ and
$\tw_1^\pm$ with mass $\sim 100-300$~GeV while $\tz_1$ is the LSP which
is dominantly higgsino-like (albeit with a non-negligible gaugino
component).  Due to the compressed spectrum amongst the various higgsino
states (typically a 10-20~GeV mass gap in models with gaugino mass
unification), their three-body decays yield only tiny visible energy
release, making them very difficult to detect at LHC.  On the other
hand, the light higgsinos should be detectable at an ILC provided
that $\sqrt{s}>2|\mu |$.

\begin{figure}[htb]
  \begin{center}
\includegraphics[width=0.6\textwidth]{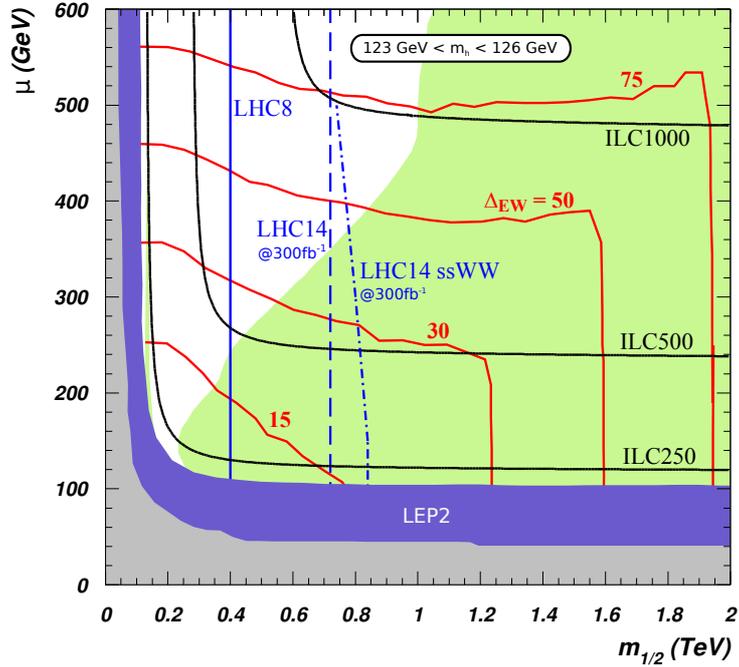}
  \end{center}
  \caption{Contours of $\Delta_{EW}=15,\ 30\ 50$ and 75 in the $m_{1/2}$ vs. $\mu$ plane for
the RNS model with parameters as shown. The blue vertical lines show the 
current reach of LHC8 and the projected reach of LHC14 with 300~fb$^{-1}$ via
gluino pair searches (dashed line) and same-sign dibosons (dot-dashed).
The reach of ILC with $\sqrt{s}=250$, 500 and 1000~TeV is also shown.
The green-shaded region has a thermal higgsino relic abundance
$\Omega_{\tilde h}h^2\le 0.12$.  
}
\label{fig:rnsplane}
\end{figure}

The situation is summarized in Fig.~\ref{fig:rnsplane} where we show the
$\mu\ vs.\ m_{1/2}$ plane from the RNS model, taking GUT-scale matter scalar masses 
$m_0=5$~TeV, $\tan\beta =15$, $A_0=-1.6 m_0$
and $m_A=1$~TeV. The Higgs boson mass $m_h\simeq 125$~GeV over the entire plane. 
%
We see that LHC8 has explored $m_{1/2}< 400$~GeV via the search for
$\tg\tg$ production.  The projected LHC14 reach with 300$^{-1}$~fb for
$\tg\tg$ production~\cite{bblt} and for same-sign diboson
production~\cite{lhcltr} extends to $m_{1/2}\sim 700-800$~GeV
(corresponding to a reach to $m_{\tg}\sim 1.8-2.1$~TeV). The naturalness
contours of $\Delta_{EW} =30$ extend well beyond LHC14 reach to
$m_{1/2}\sim 1200$~GeV.  We see that ILC600 will probe the entire
remaining parameter
space with $\Delta_{EW}< 30$, thus either discovering higgsinos or
ruling out SUSY electroweak naturalness.

\section{Acknowledgments} This research was sponsored in part by grants from the US Department of Energy


\begin{footnotesize}

\end{footnotesize}


\end{document}